# Public Finance or Public Choice? The Scholastic Political Economy As an Essentialist Synthesis

**Mohammadhosein Bahmanpour-Khalesi**

Research Fellow, Imam Sadiq (pbuh) University, Tehran, Iran. Email: m.bahmanpour@isu.ac.ir

(corresponding author)

**Mohammadjavad Sharifzadeh**

Associate Professor of Economics, Imam Sadiq (pbuh) University, Tehran, Iran. Email: sharifzadeh@isu.ac.ir

**Abstract**

Nowadays, it is thought that there are only two approaches to political economy: public finance and public choice; however, this research aims to introduce a new insight by investigating the scholastic sources. We study the relevant classic books during the thirteenth to the seventeenth centuries, and reevaluate the scholastic literature by doctrines of public finance and public choice. The findings confirm that the government is the institution for the realization of the common good according to scholastic attitude. Therefore, scholastic thinkers saw a common mission for the government based on their essentialist attitude toward human happiness. Social conflicts and lack of social consent are the product of diversification in ends and desires; hence, if the end of the humans were unified, there would be no conflict of interest. Accordingly, if the government acts according to its assigned mission, the lack of public consent is not significant. Based on the scholastic point of view this study introduce the third approach to political economy, which can be, consider an analytical synthesis among classical doctrines.

**Keywords:** Scholastic Economics, Essentialism, Political Economy, Public Finance, Public Choice, Arrow's impossibility, Human Happiness

# 1. Introduction

Although every social structure requires a special governing system, the role of the government as one of the most important socio-economic actors cannot be neglected. With the attack of the barbarians on the Western Roman Empire, the concentration of power was taken away from the central governments, and a set of tribal lives was formed under the supervision of local governments (Ganshof 1996, 4). Over time and with relative stability, this socio-economic system in its perfect form became known as Feudalism in the Middle Ages. Since the spread of insecurity and civil wars prevented the emergence of a powerful central government, in Feudalism, a central king by giving his land to vassals set up a group of independent local governments that were allied with him. While this type of governance guaranteed the security of fiefs to some extent, could also create a kind of financing for public costs through the feudal obligations to which the vassals were committed.

On the emergence of cities and central governments, feudal relations faded, and Governing institutions were extended (Pounds 2014, 432). The emergence of governments' treasury, taxes, positive laws, monetization of relations between governments and peoples, increase in the government obligations as well as public expenses were among these institutions. Since they were impressive in the humans life and interactions, their development was not important just in the practical dimensions of a socio-economic system. Therefore, they could also make many theoretical challenges for the thinkers since the middle ages. Since actions of people were considered an important factor for the humans happiness (Hirschfeld 2018, 99), every factor affecting personal and social interactions could be an analytical issue in the scholastic[1] theology. On the other hand, the interaction of divine and positive laws, which had been intensified during the Carolingian empire, was not a simple matter that scholastic thinkers could easily ignore.

Finally, the set of these factors lead the scholastic thinkers to consider the issue of the government and governance as one of the analytical topics for scholastic thinkers.

It is often assumed that the history of political economy is divided into two doctrines of public finance and public choice (Buchanan and Musgrave 1999). in this regard, most of the ideas, which are rooted in kingdom structures are attributed to the public finance approach, and ideas emphasizing modern governance aspects like democracy (Gunning 2003), liberty, individualism (Buchanan 1986), and the collective decision-making are considered as public choice. The reason for this attitude is rooted in a logical analogy arising from the definitions of these two doctrines. Nevertheless, it seems that this division about the new political economy is not a perfect view. Among the reasons for this issue, it can be traced back to the high influence of the church in government affairs and the lack of formation of centralized and high-power governments in the Middle Ages, which has prevented researchers from paying significant attention to the evaluation of the political economy literature. Accordingly, by investigating the scholastic political economy and evaluating it according to the doctrines of public finance and public choice, this study tries to introduce a third attitude, which can be considered as an essentialist insight into the relation between government and economy.

This research will be done concentrating on the realist scholastic thinkers with a phenomenological approach. In what follows, first emphasizing on happiness the role of the government in scholastic theology will be investigated. In the next step, by introducing the doctrines of public finance and public choice, the scholastic political economy will be evaluated according to them and finally, the essentialist political economy theory by focusing on the scholastic theology will be introduced.

## 2. Literature Review

Although there have been various types of research on different aspects of the scholastic economy, such as interest and usury, money, and just price, research in the field of political economy and the public sector have a small share. Among the research carried out in these two fields, most of the efforts have been dedicated to the issue of public finance and specifically justice in taxation. For example, Perdices de Blas and Revuelta López (2011) emphasized the factor of ability to pay, and Meredith (2008) the factor of utility as a core element in the scholastic tax theory.

In addition, Schwartz (2019) considered the role of scholastic thinkers in condemning the buying and selling of votes in elections. On the other hand, Urban (2014) by evaluating Saint Thomas Aquinas' theory of political economy proved the incompatibility of Aquinas's approach with the welfare state doctrine. Finally, in a study on economic liberalism in the scholastic thought, Langholm (1982) explained the role of economic ethics in limiting the boundless liberalism in the Roman laws.

As a result, the theoretical literatures demonstrate that the Works of Research conducted on the scholastic political economy show that so far no research has been done in evaluating the scholastic literature by emphasizing the distinction between the two doctrines of public finance and public choice.

## 3. The Role of the Government in the Scholastic Happiness Theory

Human happiness is one of the central dimensions of scholastic theology according to which other aspects of the scholastic literature can be interpreted. In this regard, the connection of the perfect happiness that is the vision of divine essence (Aquinas 1947, I, II, Q4, A5-6) with imperfect happiness was the special innovation of scholars that could be a foundation for the entrance of scholastic thinkers into the issue of humans' material life. In addition, it could free religion from focusing on the abstract and metaphysical aspects of the humans. Base on this attitude imperfect happiness could not bring the humans to their perfect end solely, but it could leads to some of perfection. In addition, human virtues are divine signs in the material world that could be viewed as the axis of human happiness in bodily life (Hirschfeld 2018, 99). Since virtuous persons need some material provisions, scholastic thinkers consider bodily life an important aspect of imperfect happiness (Aquinas 1947, I, II, Q4, A6) and in this regard, they started to analyze humans' bodily life and the solutions for its preservation and survival.

The analysis of the government emergence in the scholastic literature is rooted in the dissection of the human society necessity in the realization of happiness. Without considering the social life, the thinkers could not establish the social duty for the government as a natural institution. Hence, as Mariana mentioned, when there is a requirement for social life, the government could be justifiable to realize human perfect end. (Chafuen, 2003, 55).

Scholastic thinkers specified that human is naturally a social being. This inherent characteristic could be ascertained from several dimensions the most important of which was teleology. According to their teleological attitude, the thinkers believed that all beings have an end, which is their happiness (Aquinas 1947, I, II, Q1, A1-2). Wherefore, they held that social life is an important element of happiness. In this regard, Soto( [1556]1968, I:Q II:A I) stated that "[m]en who are scattered and solitary can neither teach the ignorant nor control criminals nor can they help each other to happiness by salutary advice or warnings, as can those in a community".

If virtue considered as imperfect happiness of humans in their bodily life, and the material life required social support, acting on some virtues makes social life inevitable. In this regard, Mariana ([1611] 1969, c1) expressed that nothing is more valuable than mutual charity and friendship; nevertheless, mutual charity and friendship could not be realized but in social life. In addition, scholastic thinkers maintained that humans have different talents and abilities that leads to division of labor in society. With the development of civilizations, the humans need increase. As a result, persons cannot satisfy all their needs lonely (Aquinas 1956, Q7:A17). Hereupon, this problem became another field to prove the necessity of social life.

Consequently, in the first step the thinkers ascertained the essentiality of social life for humans. In the next step, attention to the need for the government concerning the preservation and survival of social life was paid, and accordingly, the role of the government in human happiness proved. Although the scholars did not mention the details of the reasons for the failure of private mechanisms in providing the common good, they had briefly reached the fact that the realization of imperfect happiness requires conditions and facilities that are beyond the scope of private relations. They believed that the emergence of the government for supporting society toward the realization of imperfect happiness is necessary; in this regard, Suarez ([1612]2017,3:1) emphasized that a political society is essential because no household is self-sufficient and a it needs some political power since a society without this power cannot reach its end. Furthermore, he expressed that the provision of requirements that nobody can individually provide is the main duty of the government. In this regard, he added "[in] a perfect society, there must be some public power whose official duty is to consider and provide for the common good". By emphasizing the natural origin of government for preserving society, he ([1612]2017, 2:4) also emphasized that without governments society cannot be guided to a specific and unified end; therefore, social unity owes its persistence to the government.

Besides the theological approach, scholastic thinkers explained the roots of the government's emergence from an anthropological attitude. As Mariana ([1611]1969, III, 12:311) stated, at the beginning of the creation, people lived without any social order or laws; hence, nothing was important to them, but providing food and basic needs for themselves. In such a condition, households were the only factor that organized and brought them together. Over the time, the cohesion of these households weakened, so some household members migrated to nearby areas and formed a new household; further, several families created tribes together. At this age, their needs were very simple and mostly provided by nature. In such a social structure, everything belong to each other, so there was no conflict of interest, and the private property was not mattered. Nevertheless, Mariana believed that this situation could not be persistent; he insisted that over the time, people's sense of infinity led their needs to be expanded, in a way that the existing conditions did not satisfy them. From his point of view, the sense of infinity was like a double-edged sword. Although it may led to some social challenges, people's sense of infinity could increase the tendency toward perfection and the realization of a perfect society for the common good. He also added that in addition to this tendency, the ability of speech allowed neighbors to negotiate and benefit from each other's help. Finally, inspired by the unity of animals and their community around the stronger animal, society learned to attract the support of a leader against the stronger members of the community to guarantee its security. As a result, it was a foundation for the structure of the first communities and local governments.

## 4. Evaluating the Scholastic Theory of Political Economy

In this part, the evaluation of scholastic political economy theory will be discussed. For this purpose, the researchers first take a look at the doctrines of public finance and public choice, as well as their main components, and then, while rejecting its compatibility with classical approaches, the scholastic political economy theory as a third approach will be introduced.

### *4.1. A Review of Doctrines of Public Finance and Public Choice*

With the introduction of Adam Smith's "invisible hand theory" (2010 [1776], IV:293), into economic analysis, many classical economists accepted a kind of self-motivated market order as an ideal pattern for any economic system. Although after him some other economists like Malthus (1986 [1798]) and Ricardo (1996 [1817], 53) expressed some skepticism toward the automatic functioning of the market and distribution system, these criticisms could not divert the mainstream economics from this basic attitude. With the emergence of a set of practical and theoretical challenges in the self-ordered market such as asymmetric information, externalities, and the public goods (Stiglitz, 1989), the concept of market failure in economic literature has been introduced (Medema, 2007). Following its introduction, the proper foundation was created for the government to enter as the protector of self-motivated market order in the literature on political economy. In this attitude, it was expected that the government, as a social reformer, would take decisions to maximize public welfare by evaluating the common good. As a result, it will help the market to resolve its imperfection through the provision of the common good, and imposition of regulatory taxes.

Although it was thought that the introduction of the social welfare function (SWF) would help much to the interpretation of the government role in the market and to its duty in elimination of the market imperfections, the generality of the concept of welfare led to some challenges in the methodology of calculating SWF. In this regard, economists are faced with three crucial questions; first, what can be the objective nature of social welfare? Second, does public consent exist for it? Third, how can the welfare of all individuals in society be calculated to achieve a SWF for society as a whole?

However the role of the government in maximizing the SWF and minimizing market imperfections was considered the main duty of the government in the mainstream political economy for long years, a tremendous reformation in political economy was made by Arrow's impossibility theorem (1950), which made a basis for the doctrine of public choice. The Arrow's main idea was that there is no SWF to be able to aggregate the preference of all individuals simultaneously without the violation of at least one of the following properties: no dictatorship, independence of irrelevant alternatives, unrestricted domain, transitivity of preferences, positive association of social and individual values, and Pareto efficiency.

With the failure of the SWF, public choice theory was introduced as an alternative approach; however, its beginning was not restricted to the collapse of the SWF theorem. The public choice theorists considered the public finance doctrine according to a kind of anthropology, which could not be realistic in their view; they believed that the public finance theory considers the government as a benevolent person who does not pay attention to anything but the common good of its society. Nevertheless, they thought the government like other people follows just its own utility; therefore, it could not be expected that the government, like other market actors, will not seek its own interests and as a full representative, they will only pursue the public interests of their clients (Jensen and Meckling, 1994). Hence, they thought that social functions of the government are consequences of its private motivations such as gaining wealth, power, and fame; Hereupon, the government never makes decisions and choices based on the common good as it is thought in public finance approach.

Furthermore, by eliminating the paternalistic view about the government and the cultivating of an individualistic attitude toward government, the public choice doctrine could apply the microeconomics theories in analyzing the political phenomenon. In this regard, the government was considered a firm that makes political decisions with cost-benefit considerations. Additionally, the environment of political interaction can also be considered a kind of market. Public choice experts believed that political decisions are the result of a complex process, which is influenced by many actors and is not limited to a benevolent ruler. Accordingly, the collective action mechanism became the main problem in public choice theory.

In the public choice attitude toward the relations between religion and politics, all the ideologies were considered a tool to get votes for political actors. Since getting full information about the attitudes and decisions of political parties is hard for voters; therefore, they prefer to vote according to the ideologies of these parties as an essence of their attitudes. As a result, parties would try to offer ideologies that have the highest probability of getting votes (Downs, 1957).

Consequently, the theory of public choice reduced the analysis of the government and its interactions in society to the level of individual analysis, so no meaningful distinction can be made in the method of analyzing choice and individual motivations with the analysis of these elements at the social and governance level. In this regard, there was no distinction between Individual and social analysis units in political economy analysis. Accordingly, Buchanan (1986), one of the pioneers of public choice doctrine, stated that it consists of such elements as "methodological individualism", "politics as exchange", and *"homo economicus"*.

## *4.2. Introducing the Third Approach: Scholastic Political Economy*

In this section by a phenomenological attitude, the researcher will investigate the question that why the scholastic political economy could not be a public finance or public choice doctrine, and then its characteristics as a third approach will be explained.

### *4.2.1. The Impossibility of Inclusion of Scholastic Political Economy into Public Finance Approach[2]*

From the scholastic perspective, the government is a legitimate force that god grants to His servants; therefore, it has divine legitimacy (Suárez [1612] 2017, iii: IV: 5). Moreover, the government was considered a tool for the realization of social happiness; therefore, it could had a great position among members of society. The government in the scholasticism was not considered a firm, which followed its self-interest; rather, it had a public mission and responsibility to help the common good and to develop virtues in society. In this regard, Aquinas ([1267]1949,65) believes that governments have the duty "first of all, to establish a virtuous life in the multitude subject to him [government]; second, to preserve it once established, and third having preserved it, to promote its greater perfection". Furthermore, he explained the basic needs that are necessary for human happiness and emphasized that virtuous life has two preconditions governments should prepare the first and most important of which is the performance of people based on a virtuous plan. The second, which is instrumental in nature, is sufficiency of material goods that are necessary for a virtuous life.

The special mission of the government resulted in the scholastic attitude that ordinary people cannot bore this responsibility; therefore, the thinkers expected the ruler of society to be a virtuous person. In this regard, Buridan ([1349] 2020, 61)[3] and Soto ([1556]1968, I: Q II: A I) specified that a government that is not a virtuous person could not rule in a virtuous manner and protect the virtues in society. Consequently, this optimistic worldview about the nature and position of the government makes it impossible to consider it as a firm that just seeks its self-interest. Hence, scholastic political economy cannot be sorted in public choice attitudes.

*4.2.2. The Impossibility of Inclusion of Scholastic Political Economy into Public Choice Approach*

As mentioned above, the ends of the emergence of governments according to the scholastic attitude were very prominent. On this basis, the government was an instrument for realizing imperfect human happiness. Even so, did this mean complete scholastic optimism about this institution and the legitimacy of all its actions? The answer to this question was negative. Practically, the thinkers have been aware of the shortcomings of the governance institutions and the possibility of prevailing motives and interests that are not aligned with justice and the common good among bureaucrats. In this regard, attention to personal interests, rather than the common good, was always one of the concerns of scholastic thinkers about governments; for example, in this context, Buridan([1349]2020,60) specified that "the prince is sometimes very unjust, and favors his private interest". Furthermore, Mariana (1950,548), by expressing his concerns about the corruption of bureaucrats in a public position stated that "how sad it is for the republic and how hateful it is for good people to see those who enter public administration when they are penniless, grow rich and fat in public service." The lack of optimism of the scholars toward the governance institution can be found in different aspects of the scholastic political economy theory. Some of these cases are as follows:

**A. limitation of Power:** Although the government had an undeniable role in realizing human happiness, scholastic thinkers were not optimistic about the boundless expansion of political power in all different dimensions of society. In this regard, Mariana ([1609] 2002, 5) stated,

> [As], in the case of other virtues, power has definite limits, and when it goes beyond limits, power does not become stronger but, rather, becomes completely debilitated and breaks down … However, royal power increased beyond its limits is proven to degenerate into tyranny, a form of government that is not only base but weak and short-lived.

**B. Transparency of the Government:** Another dimension of pessimistic attitudes of scholastic thinkers toward government can be viewed in their emphasis on the necessity of transparency in all aspects of governance. In this regard, while condemning the rulers corruption and emphasizing the possibility of their abuse of the prestige and position, Mariana ([1609] 2002, 58-59) specified that bureaucrats must disclose all their assets before assuming responsibility and be subject to periodic supervision during their tenure like bishops.

**C. Ceasing Paying Taxes:** Another dimension of the scholastics' pessimism toward the rulers is reflected in their statement regarding the limitation of paying taxes in case of not fulfilling the duties of governance. In this regard, Aquinas ([1270] 1947,II: II: Q.62:A.7) believed that if the king takes the legal taxes but fails to fulfill its duties toward people to preserve the common good, the king will commit a sin and injustice. In this situation, the king is obliged to repay the collected taxes in case of neglecting his duties. Because "their salary is given to them in payment of their preserving justice here below".

**D. Overthrowing a Cruel and Unjust Ruler:** Perhaps one can see the extreme attitudes of the scholastic thinkers about the government in their approaches to the unjust rulers and laws. The thinkers were not stopped by ascertaining the necessity of government by natural law. According to a teleological approach, since the government was considered an instrument in realization of the imperfect happiness of humans therefore its legitimacy was dependent on it. Hereupon, there is no reason for accepting a state, which acts against its legitimate end (happiness). The thinkers believed that people not only should not obey unjust rulers but also have the right to overthrow this kind of state with pride (Buridan, [1349]2020,p.61; Suárez[1612] 2017, iii:x: 7). Additionally, some scholastic thinkers considered the right of overthrowing of the unjust government as a duty of the people. They maintained that if people do not overthrow their unjust governments, they also share the sins of the oppressive ruler (Vitoria [1528]1991, 21). In this regard, Mariana ([1611]1981, I, 6:83) with a pessimistic view about the government specified that there should always be the threat of death punishment for the rulers if they do not follow the divine and natural laws, as it could act as a deterrent for their unjust actions.

It can be concluded from all cases mentioned above that scholastic thinkers did not have an optimistic and paternalistic view about the government. They saw it as an institution that did not follow personal interests, but try to realize the common good in policies. As a result, scholastic political economy theory cannot be considered a public finance approach.

*4.3 The Scholastic Political Economy as an Essentialist Approach*
According to two previous sections, it was made explicit that the scholastic political economy cannot be considered as a public finance or public choice attitude. Now the question is how this optimism and pessimism can be interpret this optimism and pessimism in one theory?

Scholastic thinkers maintained that like other blessings, the government is a divine gift, which is given to society for the preservation and survival of social life. Even so, they believed that people would give this power to a specific ruler through a kind of social contract, so the government would be the trustee in this position (Suárez [1612] 2017, iii, iv: 5; Vitoria [1528]1991, 21). Accordingly, the government is a public representative for realizing the common good for human happiness; therefor, as long as it moves toward this end, the ruler deserve this position. Although the ruler is in charge of assigned missions and duties, this does not mean he/she is regarded as a perfect person with extraordinary abilities. Nevertheless, as the thinkers mentioned, the more virtuous the rulers of society, the better the fulfillment of their responsibility. This attitude leads the government to be seen neither as a reformer nor as a profit-seeking firm.

This attitude can only be the product of an essentialist approach to the nature of the humans and their happiness. In the scholastic literature humans were considered as a single species who have unified inherent characteristics. The more a person can actualize these characteristics, the closer he/she gets to the end that God set in her creation. In this regard, Aquinas (1947[1485], I, II, Q.1,A.7) believed that even if people choose different ends in their life, this is not due to the true distinction of the path of happiness of different human beings; nevertheless, it is due to their mistake in recognizing the real last end in life.

Furthermore, scholastic thinkers have considered an ideal pattern of the human perfection , a transcendent society, and all virtues that they could acquire; therefore, it was expected that humans would be able to take steps toward becoming a perfect being and realizing the city of God. In this regard, the harmony of ideas and attitudes about the humans resulted in a kind of social order according to the coordinated movement toward divine ideal. In this worldview, governments were considered public institutions, which have the mission of unification of all public forces in line with the divine ideal.

In the scholastic literature, governments had a special mission in realizing the common good, so if a government makes policies in the line with its assigned duties, the common good will happen without harming any member of society. Based on essentialism, despite diverse incentives, all policies of divine governments that act according to natural and divine laws would also be compatible with human essence, so no one could be damaged by those policies.[4] In this regard, the SWF will be constituted based on the inherent characteristics of the human, such that there was not any problem in public decision-making and collective actions at the macro level. Nevertheless, there is the possibility of disagreements and conflicts in strategies toward realizing divine and natural laws in society. Problems in collective decision-making or emergence of dictatorship is possible when people have different patterns for their happiness according to their desires (not inherent characteristics); however, when one consider a single end for human happiness, and the government is also in the service of realizing this supposed end, the problem of not reaching a consensus about the general policies will no longer be mattered.

Scholastic thinkers maintained that governments, despite their divine legitimacy, are independent institutions of churches (Soto [1556]1968, iv:Q iv:A I). Therefore, they belived that governments are free to use different strategies to carry out their tasks based on divine laws, and in this way, people and the church will only have a supervisory role. Although in the city of god the mission and the end of governments were assigned, this issue does not lead to the lack of dynamic prescription of different policies and patterns for the realization of happiness in society.

Consequently, since there was a specified pattern for human happiness in the scholastic literature and the governments were also obliged to make policies to realize it, this issue results in certain tasks and missions for them, and their performance can be evaluated based on it. Because there is a possibility of errors and mistakes in the actions of the rulers,[5] scholastic thinkers considered it important to evaluate their actions based on an a priori model. Therefore, while they believed in the noble mission of the government in helping human happiness, there is always a possibility of deviation of ruler deviating from the assigned duties. Furthermore, emphasizing the issue of supervision and the non-hereditary nature of state led to the more realistic scholastic political economy theory compared to public finance approaches.

## 5. Conclusion

In this research, while examining the political economy of scholastic thinkers - with an emphasis on realist scholars - from the perspective of the two doctrines of public finance and public choice, the scholastic political economy theory was introduced as a third doctrine alongside the common approaches to political economy. The results show that scholastic thinkers by using the factors of teleology and essentialism created a new approach to the political economy, which can be considered an analytical synthesis. For this purpose, they first determined the relation between the government as a social institution with human happiness, and through the teleological method, they proved the legitimacy of the government in human's social life. In the next step, based on an essentialist attitude and by focusing on the issue of virtue, they drew a unified pattern of incomplete happiness for all of human beings. In this regard, the duty of guiding society toward the common good and other virtues, helping to maintain the material survival of society through the provision of the public good and services, which is beyond the responsibility of private relations, as well as building social institutions within the framework of divine and natural laws delegated to the government. Since in an essentialist attitude, it is possible to consider a specific model for human happiness, the scholastic thinkers were able to explain the possibility of evaluating the performance of the government by defining a specific mission for them. Although the government had divine legitimacy from their point of view and the ruler needed to be a virtuous pattern for its members in society, this did not lead to the adoption of a completely optimistic attitude toward the government by the thinkers. From the scholastic perspective, rulers were considered like other members of society, so the possibility of following self-interest or making mistakes could not be deniable. Therefore, controlling the performance of the government was one of the duties of the community members in the scholastic political economy. In addition, the governments had the right to rule over the people until they take steps in line with the mission, which was settled for them.

In the public choice theory, since it is assumed that the humans have diverse interests, public decisions would have winners and losers. Additionally, this issue results in pressure and interest groups created to influence the decisions of policymakers. However, in the essentialist insights into the political economy, since the pattern of policymaking would be founded on the unified human happiness and its inherent characteristics, no one can be a loser from public policy. On the other hand, Arrow's impossibility theorem will lose its significance in an essentialist political economy, it is maintained that government policies will be conducted according to the inherent interests of the humans. Hence, even if these policies faced people's apparent objections, these objections are due to the lack of recognition of the true source of human happiness that Aquinas pointed out.

**Notes**

1. With the foundation of schools during the Carolingian empire, the thinkers who were teaching in those schools were named scholastics. They were a group of Christian teachers whose emphasis was on a philosophical approach to the explanation of ethics and divine laws and they introduce the first scientific analyzes and treaties in various dimensions of human sciences in the Middle Ages.

2. At the first glance, it may be thought that the political structure of the Middle Ages is significantly different from the current era, and since scholastic thinkers have not had a practical encounter with this political structure, it is not possible to evaluate the scholastic political economy in the framework of modern theories. However, it seems that this concern is disputed from several angles; First, many historical works of research have proven that there is no necessity for simultaneous happening of economic reality and its theoretical analysis. Therefore, a theory may not be in harmony with the economic reality of its period, but it may be useful for the analysis of economic institutions of future periods. Some examples of this issue are the introduction of the subjective theory of money by early scholastic thinkers (Lapidus and Chaplygina 2016), or the quantitative theory of money by Martin de Azpiliqueta ([1549] 2004, 279), which was compatible with subjective theory of money (Niehans 1993). On the other hand, in a phenomenological evaluation of the possibility of scholastic confrontation with modern institutions of political economy, one can be mention to Buridan ([1349] 2020, p.60) explanation of various governance models (such as monarchy and democracy) or ethics of polling (Schwartz 2019, 21), which can be a testimony to his understanding of various government structures.

3. "no one can be a prince unless he is virtuous"

4. Usually, in reality there is not any government that acts according to natural and divine laws, so we have some losers and some gainers from policies of the government. Because this policy is not according to human essence.

5. It is important to re-emphasize those mistakes and errors are meaningful for the government when there is a fixed ideal model for the happiness of humankind. In a situation where the diverse desires of people realize their different ends, talking about the deviation of the government will not have any meaning. Because every government policy will have winners and losers depending on the personal taste and happiness of the people.

# REFERENCES

Aquinas, St T. ([1270] 1947). "Summa theologica" New York: Benziger Bros.

Aquinas, St T. ([1267]1949). "On Kingship to the King of Cyprus". trans. by Gerald B" Phelan. Toronto: The Pontifical Institute of Mediaeval Studies.

Aquinas, St T. ([1503]1956). Quaestiones quodlibetales. Marietti edition: Torino/Rome.

Arrow, K. J. (1950). A Difficulty in the Concept of Social Welfare. Journal of Political Economy, 58(4), 328–346.

Azpilcueta, M.([1549] 2004). Commentary on the resolution of money. Journal of markets and morality, 7(1).

Buchanan, James M. (1986) . the Constitution of Economic Policy, Nobel Prize Lecture. Retrieved from: https://www.nobelprize.org/prizes/economic-sciences/1986/buchanan/lecture/

Buchanan James M..; & Musgrave, R. A. (1999). Public Finance and Public Choice: Two Contrasting Visions of the State. MIT Press.

Buridan, Jean. ([1349] 2020). Commentary on Aristotle's Politics. In E. W. Fuller (compiler), A Source Book on Early Monetary Thought: Writings on Money before Adam Smith (pp. 54-70). Edward Elgar.

Chafuen, A. A. (2003). *Faith and liberty: The economic thought of the late scholastics*. Lexington Books.

Chaplygina, I., & Lapidus, A. (2016). Economic thought in scholasticism. In *Handbook on the history of economic analysis volume II* (pp. 20-42). Edward Elgar Publishing.

Ganshof, F. L. (1996). *Feudalism* (Vol. 34). University of Toronto Press.

Gunning, J. P. (2003). *Understanding democracy: an introduction to public choice*. Nomad Press.

Hamilton, B. (1963). Political Thought in Sixteenth-Century Spain: A Study of the Political Ideas of Vitoria, De Soto, Suarez and Molina.

Hirschfeld, M. L. (2018). Aquinas and the Market. Harvard University Press.

Jensen, M. C., & Meckling, W. H. (1994). The nature of man. Journal of applied corporate finance, 7(2), 4-19.

Downs, A.(1957). An economic theory of political action in a democracy. *Journal of political economy*, *65*(2), 135-150.

Langholm, O.(1982). Economic freedom in scholastic thought. *History of Political Economy*, *14*(2), 260-283.

Malthus, T. R. (1986). An essay on the principle of population (1798). *The Works of Thomas Robert Malthus, London, Pickering & Chatto Publishers*, *1*, 1-139.

Mariana, J. (1950). Del Rey y de la Institución Real, in Biblioteca de Autores Españoles, Rivadeneya, vol. 31 (Madrid: Editions Atlas,[1439] 1950)

Mariana, J. ([1611]1969). De Rege et regis institutione Libri tres: Eiusdem de ponderibus & mensuris Liber. Germany: Aubrius, .

Mariana, Juane de .(1981). La dignidad real y la educación del rey (De rege et regis institutione), ed. L. Sanchez Agesta, Centro de Estudios Constitucionales, Madrid.

Mariana, J.( [1609] 2002). "A Treatise on the Alteration of Money". *Journal of Markets & Morality, 5(2)*: 523–593.

Medema, S. G. (2007). The hesitant hand: Mill, Sidgwick, and the evolution of the theory of market failure. History of Political Economy, 39(3), 331-358.

Meredith, C. T. (2008). The ethical basis for taxation in the thought of Thomas Aquinas. Journal of Markets & Morality, 11(1).

Niehans, J. (1993). "A Reassessment of Scholastic Monetary Theory". *Journal of the History of Economic Thought*. 15(2): 229-248.

Perdices de Blas, L., & Revuelta López, J. (2011). Markets and taxation: modern taxation principles and the school of Salamanca. *ESIC Market Economic and Business Journal*, *138*, 91-116.

Pounds, Norman John Greville. (2014). An economic history of medieval Europe. Routledge.

Ricardo, D. (1996). Principles of Political Economy and Taxation. United States: Prometheus Books.

Smith, A. ( [1776] 2010).The Wealth of Nations. United Kingdom: Harriman House.

Soto, D de. ([1556]1968), De iustitia et iure . Madrid: Instituto de Estúdios Políticos.

Stiglitz, J. E. (1989). Markets, market failures, and development. The American economic review, 79(2), 197-203.

Suárez, F. ([1612] 2017). Tractatus de Legibus AC Deo Legislatore in Decem Libros Distributus (Classic Reprint). United States: Fb&c Limited.

Schwartz, D. (2019). The Political Morality of the Late Scholastics: Civic Life, War and Conscience. Cambridge University Press.

Urban, Konrad Edward.(2014). Aquinas and the Welfare State. .retrieved from: http://2014.austriancenter.com/wp-content/uploads/2014/01/Aquinas-and-the-Welfare-State.pdf

De Vitoria, F. (1991). Vitoria: political writings. Cambridge University Press.